# John Dollond's Prismatic Experiments: The Beginning of a Long Way

Igor Nesterenko

*Budker Institute of Nuclear Physics, Novosibirsk, 630090, RUSSIA*
*Observatory of Novosibirsk State University*
(corresponding author, e-mail: in4astro@gmail.com)

**Annotation**

The English optician John Dollond (1706-1761) performed a number of experiments with prisms of various transparent materials and concluded that an achromatic lens could exist. At that time, the dominant opinion about light refraction denied this possibility. This work presents the computer simulations of these experiments. Simulations allowed us to indicate the boundaries on an optical parameters of glasses, which could be used by J. Dollond in his prism experiments.

**Introduction**

This work is entirely based on the account of John Dollond [1]. His work is interesting mainly because it was the first [1] publication in which the statement of Isaac Newton about impossibility of white light refraction without dispersion – without decomposition into monochromatic components, was experimentally disproved. In this work J. Dollond demonstrated the possibility of creating an achromatic lens. There he claimed to have found a solution for simultaneous elimination of spherical and chromatic aberrations in a lens consisting of two different type glasses. Unfortunately, the details of method by eliminating spherical aberration at achromaticity conservation were not disclosed either in this account nor in the patent [2] obtained about two months before presentation to the Royal Society. Instead, the J. Dollond & Son optical shop began selling telescopes with achromatic lenses which had a properties declared in the patent.

Information on early achromatic lenses from the late 50s and early 60s of XVIII century can be found in works [2] and [3]. The last work gives not only an optical design of early achromatic lenses, but also provide the optical parameters of glasses from which they were made. These glass data [3] were used at the numerical reconstruction of J. Dollond's prism experiments. In Appendix I, the optical parameters of old glasses are given. Reconstruction of prism experiments was carried out with the help of Zemax-EE software.

Detailed consideration of the account was required to obtain most reliable information about glasses of Lomonosov's telescope, which he used during the observations of Venus transit in 1761. At that time, he discovered Venusian atmosphere. It is known that Lomonosov used refractor-achromat made in Dollond's optical shop [4].

By the mid-XVIII century, the English glass industry had already produced of various glass types which necessary for manufacturing of achromats – flint and crown. However, the list of glasses, especially with optical quality, was very limited. The quality and the main optical parameters of glasses – refractive index and dispersion value (or Abbe number) had a significant variation for a different melting batches. Therefore, the information on glasses used by J. Dollond for prism experiments can be considered as a good starting point at choice of modern glasses similar to those which were used in manufacture of the achromatic doublet in Lomonosov telescope.

---

[1] During the trial in 1764 in which considering petition of 35 London opticians against to the Dollond's patent for an achromatic lens, it was recognized that the first achromatic lens had been made by Chester Moor Hall between 1730 and 1750. However, Hall did not publish his research anywhere, though in court one of Hall's achromatic lenses was demonstrated.

[2] British Patent No.721 of April 19, 1758.

[3] Four achromatic lenses are described in work [3]:
No.1 – 'J. Linnell London', Eng18_F2 (flint) and Eng18_C2 (crown);
No.2 – 'Ayscough London', glass Eng18_F3 (flint) and Eng18_C3 (crown);
No.3 – 'Dollond London', glass Eng18_F1 (flint) and Eng18_C1 (crown);
No.4 – 'Dollond London', glass Eng18_C4 (crown) and Eng18_F4 (flint).

**Three steps to achromatic lens**

"An Account of Some Experiments Concerning the Different Refrangibility of Light" by J. Dollond, was presented by James Short to the Royal Society on June 8, 1758. This account contains a description of three experiments with prisms making from different transparent materials that were carried out during 1757 and the first half of 1758. The first experiment with the water-glass prism (Fig.1) in all details repeated Newton's experiment [4], carried out 90 years before. However, the experiment of J. Dollond gave exactly opposite result. Firstly, the objects seen without refraction through the water-glass prism appeared to be painted by prismatic colors as if they were seen only through the glass prism. Secondly, his experiment showed the possibility of refraction of incident white light without dispersion – without decomposition into monochromatic components, i.e. the principal possibility of creating achromatic lens.

Figure 1. Water-glass prism.

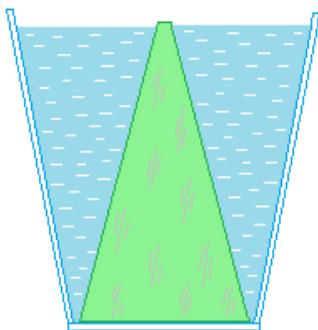

The second experiment was made in a similar geometry like the first, but the apex angle of glass prism was essentially reduced, also apparently another glass type was used. In this experiment an attempt was made to obtain quantitative conditions for the parameters of achromatic lens consisting of two glass lenses and water between them [5]. Though these experiments did not lead to a working prototype of achromatic lens, they did demonstrate the principle feasibility of such a lens.

The third experiment was carried out with glass prisms made from different types of glasses and apparently in an improved observation scheme which provided more accurate measurements. Prior to this experiment various glasses were tested according to their dispersion values. J. Dollond was looking for a pair of glasses that had the maximum dispersion difference [6]. By this time, he already understood the reason why the water-glass lens was unsuccessful. This was due to the large difference in refractive powers between glass and water (refractive indices 1.560 and 1.334 respectively). Consequently, relatively short radii of curvature for lens surfaces were required, which lead to growth of geometric aberrations (especially spherical). Thus, for a good achromatic lens, it was necessary to find two glasses with a minimum difference in refractive powers and a maximum difference in dispersions (or Abbe numbers) between them.

Simultaneously with the search for a suitable pair of glasses, theoretical investigation was carried out on a method for eliminating of geometric aberrations at conservation of achromaticity – reciprocal compensation of dispersions for two lenses.

---

[4] Isaac Newton, Opticks, Book I. Part II. Prop.3. Experiment VIII.

[5] *Having, about the beginning of the year 1757, tried these experiments, I soon after set about grinding telescopic object-glasses upon the new principles of refraction, which object-glasses were compounded of two spherical glasses with water between them. These glasses I had the satisfaction to find, as I had expected, free from errors arising from the different refrangibility of light: for the refractions, by which the rays were brought to a focus, were every-where the differences between two contrary refractions, in the same manner, and in the same propositions, as in the experiment with the wedges. However, the images formed at the foci of these object-glasses were still very far from being so distinct as might have been expected from the removal of so great a disturbance; and yet it was not very difficult to guess at the reason, when I considered, that the radii of the spherical surfaces of those glasses were required to be so short, in order to make the refractions in required propositions, that they must produce aberrations, or errors, in the image, as great, or greater than those from the different refrangibility of light.*, pp.738-739 [1].

[6] *I discovered a difference, far beyond my hopes, in the refractive qualities of different kinds of glass, with respect to their divergency of colors. Yellow or straw-colored foreign sort, commonly called Venice glass, and the English crown glass, are very near alike in that respect, tho' in general the crown glass seems to diverge the light rather the least of the two. The common plate glass made in England diverges more; and the white crystal or flint English glass, as it is called, most of all. It was not my business to examine into the particular qualities of every kind of glass that I could come at, ... but to fix upon such two sorts as their difference was the greatest; which I soon found to be the crown, and the white flint or crystal.*, p.740 [1].

The method was found [7] and soon it was patented. The details of calculation method have not been published anywhere. Apparently, the Dollond's method was based on the concept of lens and wedge equivalence when they refract light.[8] The only difference, that any lens is a wedge with a variable apex angle depending on the distance to its axis.

**Experiment 1**

*I diminished or increased the angle between the glass plates, till I found the two contrary refractions to be equal; which I discovered by viewing an object thro' this double prism; which, when it appeared neither raised nor depressed, I was satisfied, that the refractions were equal, and that the emergent rays were parallel to the incident.*
*Now, according to the prevailing opinion, the object should be have appeared thro' this double prism quite of its natural color; for if the difference of refrangibility had been equal in two equal refractions, they would have rectified each other: but the experiment fully proved the fallacy of this received opinion, by shewing the divergency of the light by the prism to be almost double of that by the water* [prism]*; for the object, tho' not at all refracted, was yet as much infected with prismatic colors, as if it had been seen thro' a glass wedge only, whose refracting angle was near 30°. I plainly saw then, that if the refracting angle of the water-vessel could have admitted of a sufficient increase the divergency of the colored rays would have been greatly diminished, or intirely rectified; and there would have been a very great refraction without color, as now I had a great discoloring without refraction: but the inconveniency of so large an angle, as that of the vessel must have been, to bring the light to an equal divergency with that of the glass prism, whose angle was about 60°.* pp.735-737 [1]

This experiment completely repeats Newton's 8th experiment from the book "Optiks". However, Dollond's result is qualitatively different. This difference was explained by Peter Dollond in his work [5], J. Dollond eldest son and his companion in optical shop.
The key difference in Newton's and Dollond's experiments was in using of prisms which were made from different glass types. Newton probably chose Venetian glass for his prisms. It was slightly straw-colored. In the 60s of XVII century, quality of this glass was better than of English crown, which had a distinctive green tint due to iron oxide. At that time, flint (colorless glass) was not yet produced by the English glass industry. To check his hypothesis, Peter Dollond made a similar prism as Newton from Venetian glass and obtained the same result. The Fig.2 shows what approximately Newton and Peter Dollond could see through such water-glass prism combination.
After 90 years, J. Dollond had great opportunities in choosing the type of glass for his prism. At studying the coloration of objects visible through prisms, it is desirable to use colorless transparent materials, so the choice of colorless flint in the 50's of XVIII century was quite natural.

---

[7] *Having thus got rid of the principal cause of the imperfection of refracting telescopes* [chromatism], *... before I found, that the removal of one impediment had introduced another equally detrimental (the same as I had before found in two glasses with water between them): for the two glasses, that were to be combined together, were the segments of very deep spheres; and therefore the aberrations from the spherical surfaces became very considerable, and greatly disturbed the distinctness of the image... the surfaces of spherical glasses admit of great variation, tho' the focal distance be limited, and that by these variations their aberrations may be made more or less, almost at pleasure; I plainly saw the possibility of making the aberrations of any two glasses equal; and as in this case the refractions of the two glasses were contrary to each other, their aberrations, being equal, would intirely vanish. And thus, at last, I obtained a perfect theory for making object-glasses, to the apertures of which I could scarce conceive any limits: for if the practice could come up to the theory, they must certainly admit of very extensive ones, and of course bear very great magnifying powers.* pp.741-742 [1].
[8] *... as the refractions of spherical glasses are in an inverse ratio of their focal distances; it follows, that the focal distances of the two glasses should be inversely as the ratio's of the refractions of the wedges: for being thus proportioned, every ray of light that passes, thro' this combined glass, at whatever distance it may pass from its axe, will constantly be refracted, by the difference between two contrary refractions, in the proportion required; and therefore the different refrangibility of the light will be intirely removed.* p.741 [1]

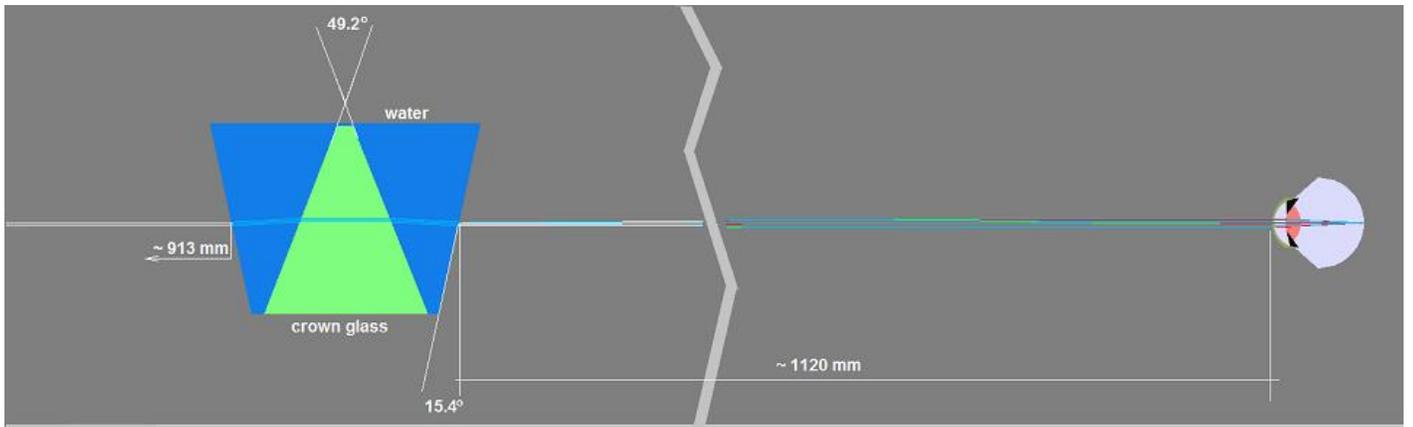

Figure 2. Image of the text describing experiment #8 from Newton's book "Optiks".
A glass prism made from a crown which has optical parameters close to Venetian glass. The letters outside the green circle are slightly colored due to non-zero refraction angle. The prism from Venetian glass with apex angle 49.2° gives a refraction angle about 30°.

In experiment #1 a glass prism refracted light at angle of almost 30°. Here it should be mentioned that under "prism" we mean an isosceles wedge, and under "refraction angle" is the smallest angle of refraction[9]. This information is not sufficient to identify the refractive index of glass prism, because it is also necessary to know its apex angle, which is not specified. However, it is additionally known for compensation of the glass prism dispersion the apex angle of water-prism should be about 60°. Again this information does not give a direct indication on the glass type, but a choice is significantly narrowed. The Fig.3 shows a map of modern glasses on the "refractive index – Abbe number" diagram[10]. At the left edge, the map is limited by water. Because the experiment description states that for water-glass prism with zero refraction, the colors of objects were as if they were visible through a glass wedge. In other words, the glass dispersion was larger (lower Abbe number) than for water.

The green strip covers the glasses which fit to the description of experiment #1. As we can see, inside the band is located the flint from Joseph Linnell achromatic lens [3].

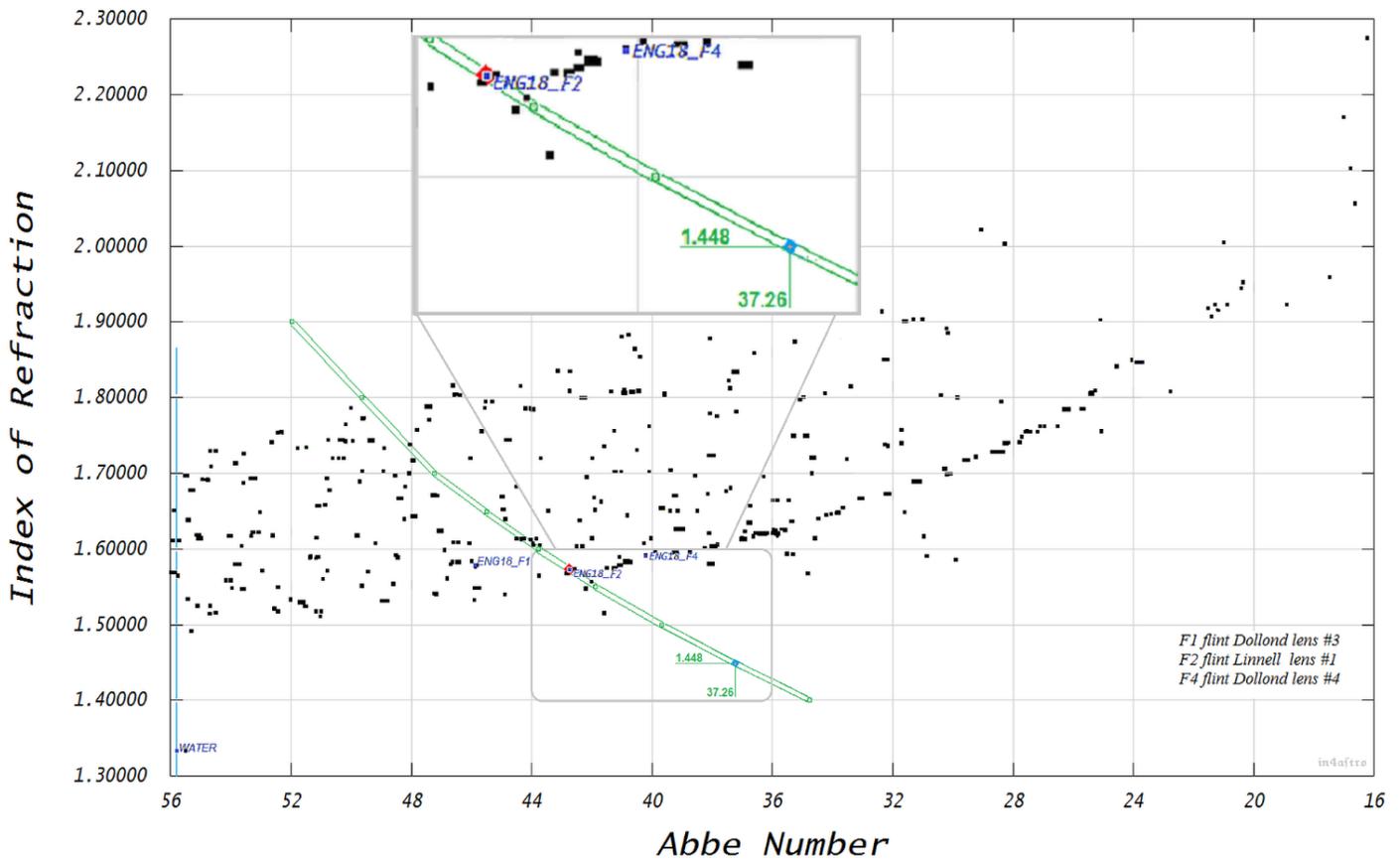

Figure 3. Map of modern glasses (black rectangles) on the diagram "refractive index – Abbe number" from catalogues: Schott, Ohara, USSR, special glasses and English glasses of XVIII century from work [3]. The green stripe covers the glasses satisfying the conditions of experiment #1. The upper edge of the band corresponds to the apex angle of water-prism (at compensated dispersion) 60.4° and the lower edge 59.6°. The red color indicates location on the diagram of Eng18_F2 flint from the Linnell lens. The numbers (1.448 and 37.26) indicate on the location of hypothetical glass satisfying the "doubled dispersion" condition (details in text).

---

[9] The smallest angle of refraction occurs when light propagates parallel to the prism base.
$n \cdot \mathrm{Sin}[\alpha/2] = \mathrm{Sin}[(\theta+\alpha)/2]$, here $\alpha$ is the prism apex angle, $\theta$ is the prism refractive angle. This relation is used for calculation of refractive index and this method is often called Newtonian.

[10] All "refractive index – Abbe number" diagrams are given for the yellow sodium line. Abbe number is equal to ratio $(n_d - 1)/(n_F - n_C)$, here n – refractive index and the letter indicates wavelength: d – 0.5876 μm, F – 0.4861 μm and C – 0.6563 μm at which it was measured.

The Fig.4 shows an image of text which J. Dollond could see at zero refraction angle of the water-glass prism with Eng18_F2 glass[11] (flint from Linnell lens), and the Fig.5 shows all the same, but seen through a water-glass prism with compensated dispersion. In the last case, the refraction angle of axial beam is about 10.5°[12]. The glass prism (F2 flint) with apex angle 46° has refraction angle 30°. To compensate the dispersion[13] of such prism the water prism with apex angle 60.1° is required. This is in good agreement with the description of experiment #1. However, there is one essential difference:

J. Dollond reported that the dispersion of water-glass prism at zero refraction angle had *double dispersion* to the water prism with the same apex angle. But this does not match properties of a glass prism made from F2 flint. On the contrary, in this case a dispersion of the water-glass prism compared with the water prism (without of the glass prism) is 2.1 time less. The dispersion ratio was probably recorded as a fraction "1/2" in a workbook, but for some unknown reason the first half – "1/" was lost from the account. On the Fig.3 numbers 1.448 and 37.26 (refractive index and Abbe number, respectively) show the position of hypothetical glass on the diagram which would give the reported *double dispersion* a water-glass prism at zero refraction angle to a simple water prism with the same apex angle. It is easy to see that such glass does not exist.

---

[11] The "Eng18_" will hereafter be omitted in the names of XVIII century English glasses.

[12] Axial beam always means green light with a wavelength of 0.555 μm which corresponding to the maximum spectral sensitivity of human eye.

[13] Dispersion compensation refers to the convergence of blue (0.4861μm) and red (0.6394μm) rays. At these wavelengths, eye sensitivity is about 18% of maximum. If the glass data should be plotted, the refractive indices were recalculated with using Conrady formula to the wavelengths indicated in footnote 10.

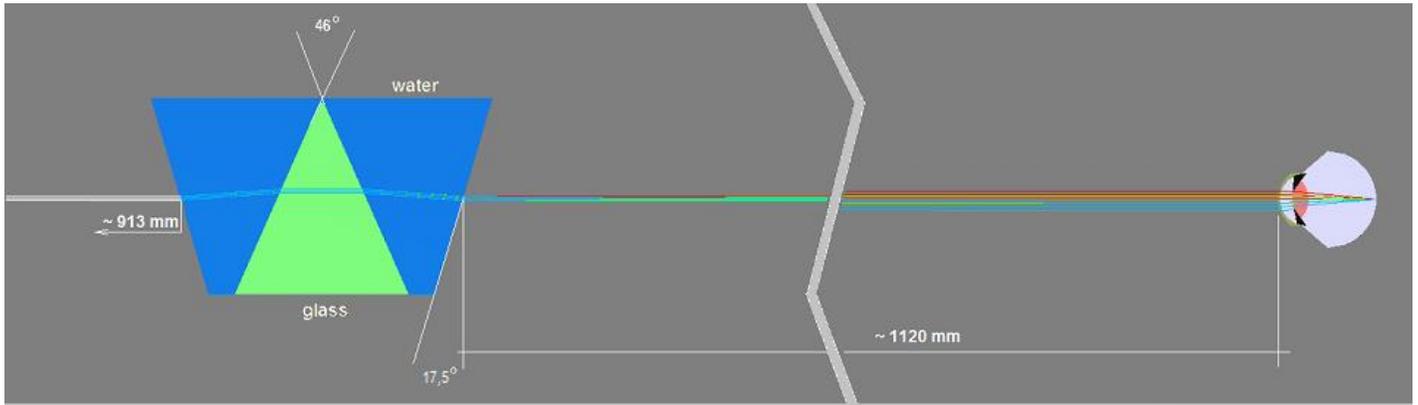

Figure 4. Image of the text describing experiment #8 from the Newton's book "Optiks".
The green prism with apex angle 46° is F2 flint glass from the Linnell lens. The water prism with apex angle 35° is required for refraction compensation of the glass prism.

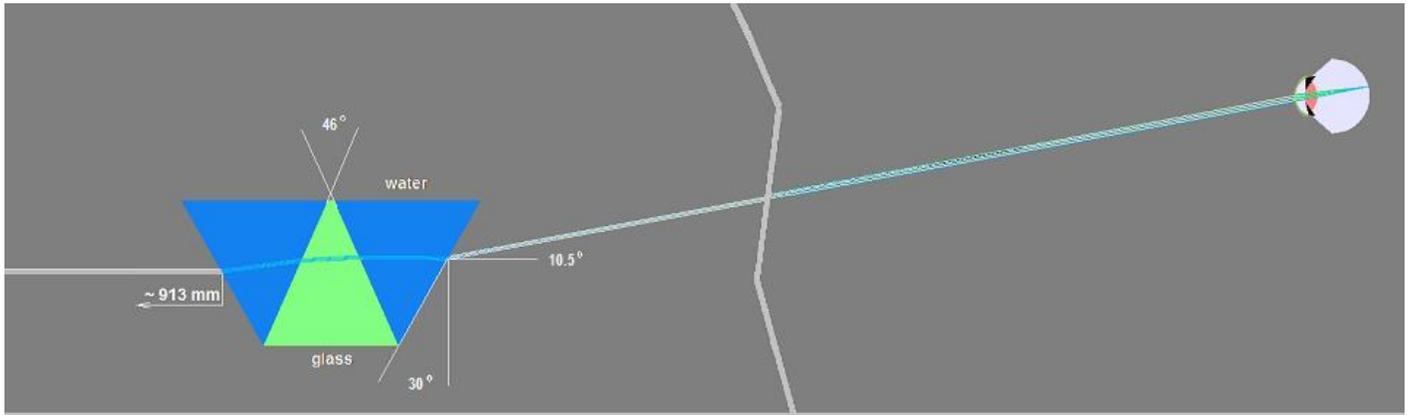
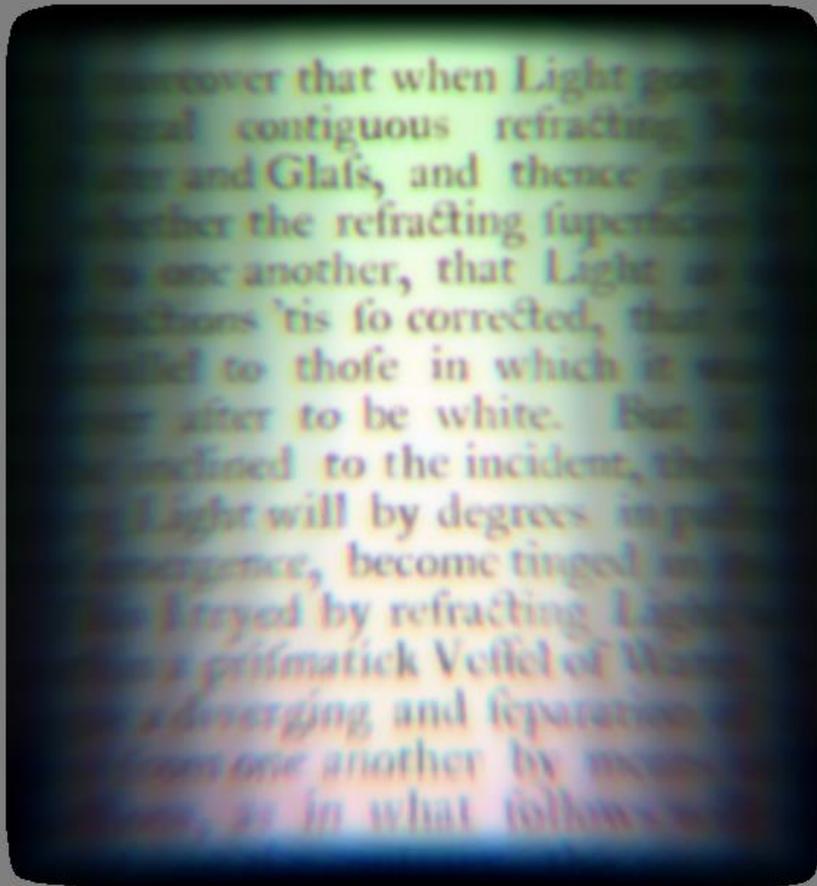

Figure 5: Image of the text describing experiment #8 from the Newton's book "Optiks".
The green prism is a F2 flint glass from the Linnell lens. The water prism with apex angle 60.1° is required to compensate a dispersion of glass prism. The total refraction angle is about 10.5°. As in the case of Newton's experiment, dispersion compensation takes place only for rays near axis (at the picture center). For off-axis rays the condition of achromatic compensation is violated due to glass and water prisms have very large apex angles.

**Experiment #2**

*I ground a wedge of common plate glass to an angle of somewhat less than 9°, which refracted the mean rays [light with a wavelength 0.55 μm corresponding to the maximum sensitivity of eye] about 5°. I then made a wedge-like vessel, as in the former experiment, and filling it with water, managed it so, that it refracted equally with the glass wedge; or, in other words, the difference of their refractions was nothing; and object viewed thro' them appeared neither raised nor depressed. This was done with an intent to observe the same thing over again in these small angles, which I had seen in the prism: and it appeared indeed the same in proportion, or as near as I could judge; for notwithstanding the refractions were here also equal, yet the divergency of the colors by the glass was vastly greater than that by the water; for objects seen by these two refractions were very much discolored. Now this was a demonstration, that the divergency of the light, by the different refrangibility, was far from being equal in these two refractions. I also saw, from position of the colors, that the excess of divergency was in the glass; so that I increased the angle of the water-wedge ; by different trials, till the divergency of the light by the water* [prism] *was equal to that by the glass* [prism], *that is till the object, tho' considerably refracted, by the excess of the refraction of the water, appeared nevertheless quite free from any colors proceeding from the different refrangibility of light; and, as near as I could then measure, the refraction by the water was about 5/4 of that by the glass.*
*Indeed I was not very exact in taking the measures, because my business was not at that time about proportion, so much as to shew, that the divergency of the colors, by different substances, was by no means in proportion to the refractions; and that there was a possibility of refraction without any divergency of the light at all.*
pp.737-738 [1]

The description of this experiment is more detailed and allows to significantly limit the area of suitable glasses. Since in the description of prism parameters not exact values of angles are given, but only approximate ones, the area should be extended additionally at least by measurement error of angles. The accuracy of angle measurements was assumed ±0.1°. The angle of glass prism was slightly less than 9°, so it was taken as 8.9°±0.1° and the angle of refraction as 5.0°±0.1°. From these data it follows that the refractive index is 1.560±0.013 [14].

Two prisms are achromatic to each other (red and blue rays have the same total refractive angle) if the ratio of their refractive angles is equal to the ratio of their Abbe numbers, i.e. $v_w / v_g = \theta_w / \theta_g$, here $v$ – Abbe number, $\theta$ is the prism refractive angle, and the indices denote water (w) and glass (g). Pure water has an Abbe number of 55.79 and since the ratio of the refractive angles of the water prism to the glass prism was about 5/4 (or 1.25±0.03). It means the used glass has an Abbe number about 45±1.

The Fig.6 shows a "refractive index – Abbe number" diagram. The green rectangle covers the area where the suitable glass with 68% probability should be located. The orange one covers the area where glass should be located with 95% probability. The most likely candidate from a list of modern glasses is the glass S-TIL1 from Ohara's catalog (or LLF1 Schott's analogue). J Dollond used a plate glass in this experiment. At that time, under "plate glass" was meant glass which was placed between crown and flint glasses by optical parameters, but closer to the latter. In the production of plate glass, lead oxide was also used, as in a flint glass, but in smaller quantities. Therefore, the plate glass always had a smaller refraction index and less dispersion (Abbe number is higher) than at flint glass.

The Fig.7 shows an image of text that J. Dollond could see at zero angle of refraction of a water-glass prism with S-TIL1 glass (Ohara's catalogue). In this case, the apex angle of water prism is needed about 5.8°.

The Fig. 8 shows the same, but seen through a water-glass prism with compensated dispersion. In this case, the water prism with apex angle 9.1° is required. At the same time, the water-glass prism gives the total

---
[14] The refractive index is determined from the formula: n(α,θ)·Sin[α/2] = Sin[(θ+α)/2], by the known refractive angle and apex angle. The error is calculated by the formula:

$$\sigma_n = \sqrt{\left(\frac{\partial n}{\partial \alpha} \cdot \sigma_\alpha\right)^2 + \left(\frac{\partial n}{\partial \theta} \cdot \sigma_\theta\right)^2}$$

refractive angle for axial beam about 1.1°. The achromatic effect at such angles is less noticeable than in experiment #1, but it is still visible.

The refraction angle of S-TIL1 glass prism with an apex angle 8.9° is 4.9°±0.1°. The refraction angle of achromatic to it water prism with an apex angle 9.1° is 6.1°±0.1°. The ratio of refractive angles of the water prism to the glass prism is 1.24±0.03, which is in good agreement with ratio 5/4 from the experiment #2 description.

In addition, it should be emphasized that three out of four old English flint glasses from work [3] were covered by the orange area which limits of 95% probability.

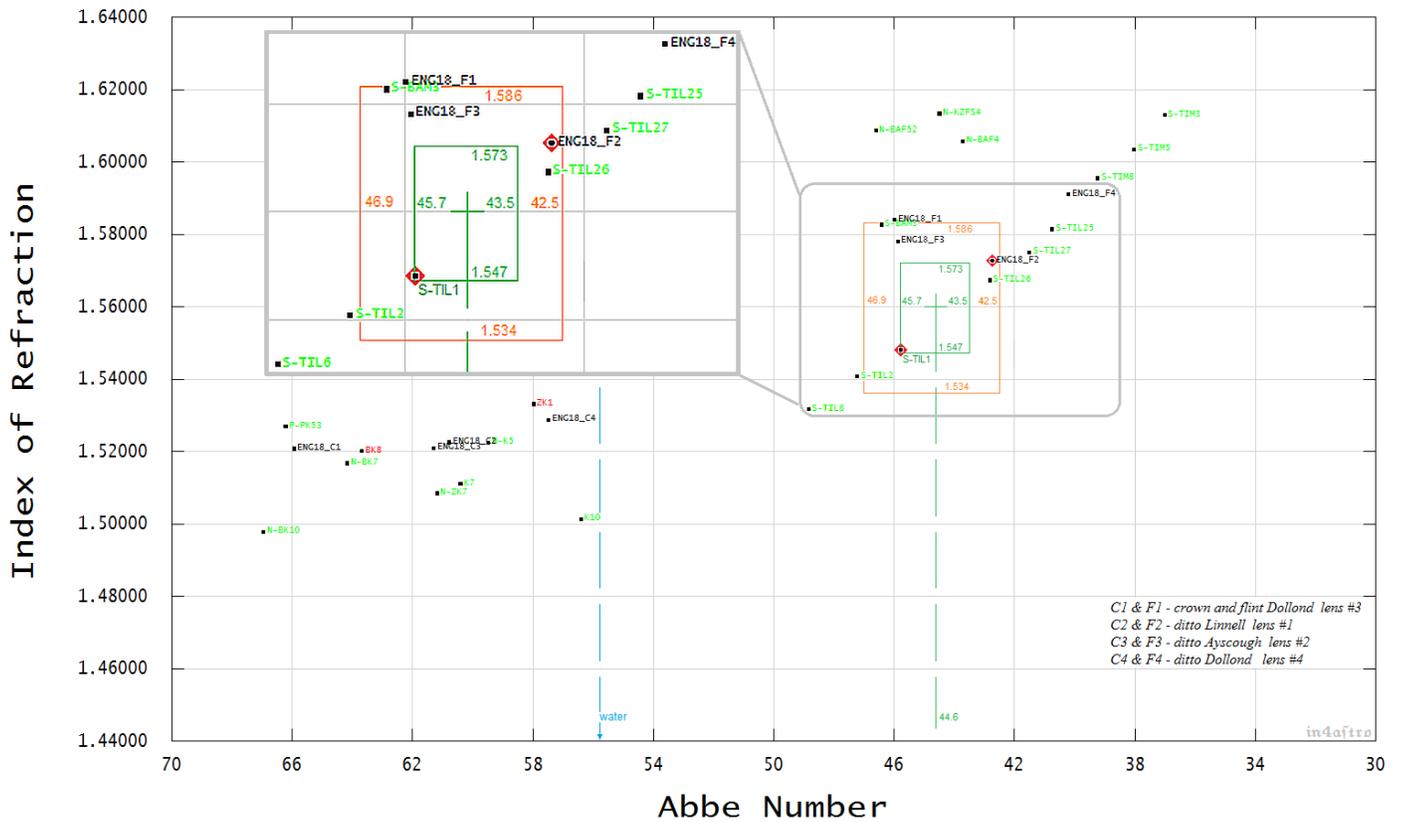

Figure 6. Diagram "refraction index – Abbe number" from catalogues: Schott, Ohara (partially) and English glasses of XVIII century. Green and orange rectangles cover areas where with a probability of 68% and 95%, respectively, should be placed a glass that satisfy the conditions of experiment #2. The values near rectangle edges indicate on refractive indices and Abbe numbers of selected areas.

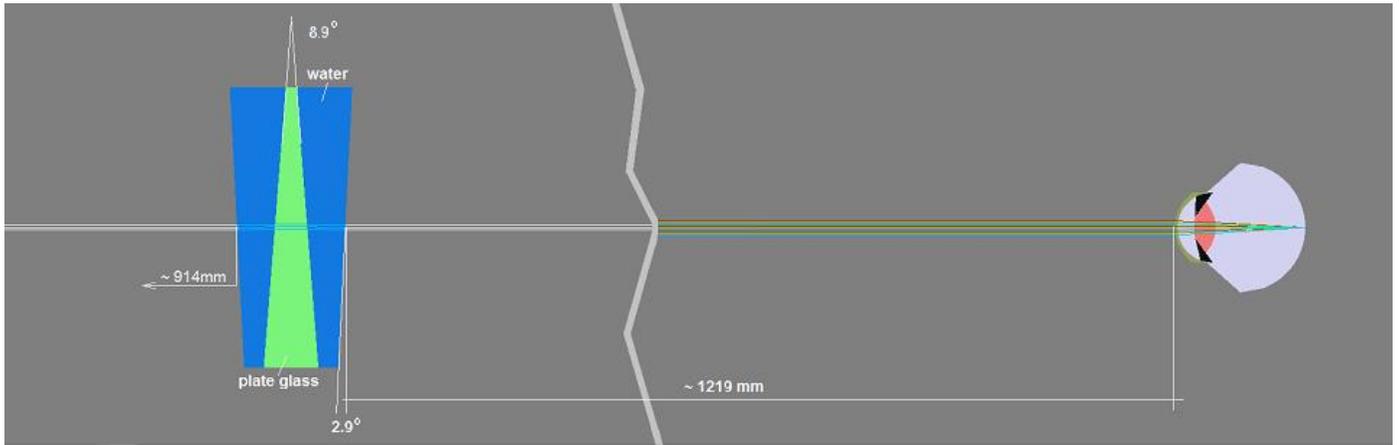

## EXPER. VIII.

I found moreover that when Light goes out of Air through feveral contiguous refracting Mediums as through Water and Glafs, and thence goes out again into Air, whether the refracting fuperficies be parallel or inclined to one another, that Light as often as by contrary refractions 'tis fo corrected, that it emergeth in lines parallel to thofe in which it was incident, continues ever after to be white. But if the emergent rays be inclined to the incident, the whitenefs of the emerging Light will by degrees in paffing on from the place of emergence, become tinged in its edges with Colours. This I tryed by refracting Light with Prifms of Glafs within a prifmatick Veffel of Water. Now thofe Colours argue a diverging and feparation of the heterogeneous rays from one another by means of their unequal refractions, as in what follows will more fully appear. And, on the contrary, the permanent whitenefs argues, that in like incidences of the rays there is no fuch feparation of the emerging rays, and by confequence no inequality of their whole refractions. Whence I feem to gether the two following Theorems.

Figure 7: Image of the text describing experiment #8 from Newton's book "Optiks".
The green prism is S-TIL1 glass (Ohara). The water prism with apex angle 5.8° is required to compensate for the refraction of glass prism.

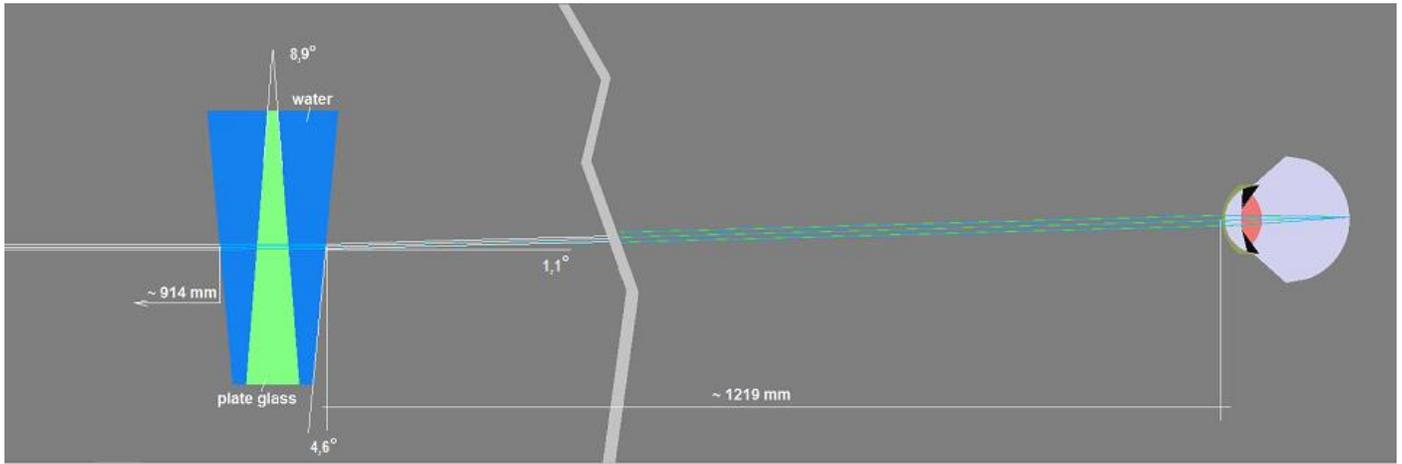

Figure 8: Image of the text describing experiment #8 from Newton's book "Optiks".
The water prism with apex angle 9.1° is required to compensate chromaticity of the S-TIL1glass prism. The total refractive angle of the water-glass prism is about 1.1°

**Experiment #3**

*And yet, as these experiments clearly proved, that different substances diverged the light very differently, in proportion to the refraction; I began to suspect, that such variety might possibly be found in different sorts of glass, ...yet I did not expect to meet with a difference sufficient to give room for any great improvement of telescopes; so that it was not till the latter end of the year [1757] that I undertook it, when my first trials convinced me that this business really deserved my utmost attention and application.*
*I therefore ground a wedge of white flint of about 25°, and another of crown of about 29°, which refracted nearly alike; but their divergency of the colors was very different. I then ground several others of crown to different angles [wedges], till I got one, which was equal, with respect to the divergency of the light, to that in the white flint: for when they were put together so as to refract in contrary directions, the refracted light was intirely free from color. Then measuring the refractions of each wedge, I found that of the white glass to be to that of the crown nearly as 2 to 3; and this proportion would hold very nearly in all small angles [wedges].*
pp.739-740 [1]

In the description of this experiment, J. Dollond again has specified incomplete information about parameters of glass prisms. As in experiment #1, it is possible to talk only about a possible variants of used glass pairs. In the first case the description was even more definite, since the optical parameters one of medium were known. From the fact, that the apex angles of flint and crown prisms were specified, when compensation of their refractive angles occurs, can be established only one-to-one mapping between flint and crown glasses without possibility of exactly determining what glass pair was really used. Also the apex angles of prisms were rounded to whole degrees. Therefore, we can suppose that the apex angle of the flint prism could be from 24.6° to 25.4° and of the crown prism from 28.6° to 29.4°. In addition, for achromatic refraction the ratio of the refraction angles of flint prism to crown prism should be about 2/3. As before let's suppose that the measurement accuracy of any angles (prism or refraction angle) in the experiment was ±0.1°. In this case, the achromatic angle ratio with tacking into account the measurement errors was 0.667±0.005. The indicated accuracies of optical parameters make it possible to establish the regions on "refractive index – Abbe number" diagram where the glass pairs for a specific crown or flint should be chosen to satisfy the description of given experiment.

The Fig. 9 shows these matched regions. For example, if we consider that the apex angles of both prisms were exactly 25° and 29° and the achromatic ratio is 2/3, then for the four old English crowns we should choose flints from the yellow regions were marked C1...C4 respectively to the name of crown glasses. And vice versa, if the old flints are chosen, they correspond to crowns from the blue regions were marked in the same manner. In both cases, the matching regions for glass pairs are placed either lower of 1.50 line for crowns, or higher of 1.60 line for flints, than the set of old glasses from work [3]. It means that the apex angles of prisms are indeed rounded to an integer value.

If we increase the flint prism angle to 25.4° and decrease the crown prism angle to 28.6°, regions for the paired flints (marked in orange) are shifted down and regions for the paired crowns (marked in green) are shifted up. The ratio of achromatic refractive angles did not change and amounted to 2/3. In this case, the pairs from the set of old glasses are already available. These are C2–F4 and C3–F4 pairs. For the pair C1 and F1 the ratio of achromatic angles should be additionally changed up to 0.7 to comply the conditions of experiment #3.

The F2 flint, which was admirably suited to the conditions of experiment #1, has no matching pair in experiment #3 from the list of old crowns. Apparently, J. Dollond in this experiment used another flint. Two facts push to such conclusion:
1) he informed in the account that the water-glass prism from the first experiment was kept by him for possible future demonstration[15];

---

[15] *N.B. This experiment will be readily perceived to be the same as which Sir Isaac Newton mentions; but how it comes to differ so very remarkably in the result, I shall not take upon me to account for; but will only add, that I used all possible precaution and care in the process, and that keep the apparatus by me to evince the truth of what I write, whenever I may be properly required so to do.* p.736 [1]

2) he noted that glasses (especially flints) had significant variation in optical properties from one melting batch to another.

Nevertheless, if Dollond's optical shop had sufficient supply of flint glass from one melting and in experiments #1 and #3 the same glass was used, then it should be placed on intersection of the orange region corresponding to C1 crown (marked in "2/3") and the light green strip. However, this region of intersection is located on the diagram, where there are no modern glasses. Maybe sometimes flint glasses of the XVIII century were accidentally dropped into this area, it is unknown.

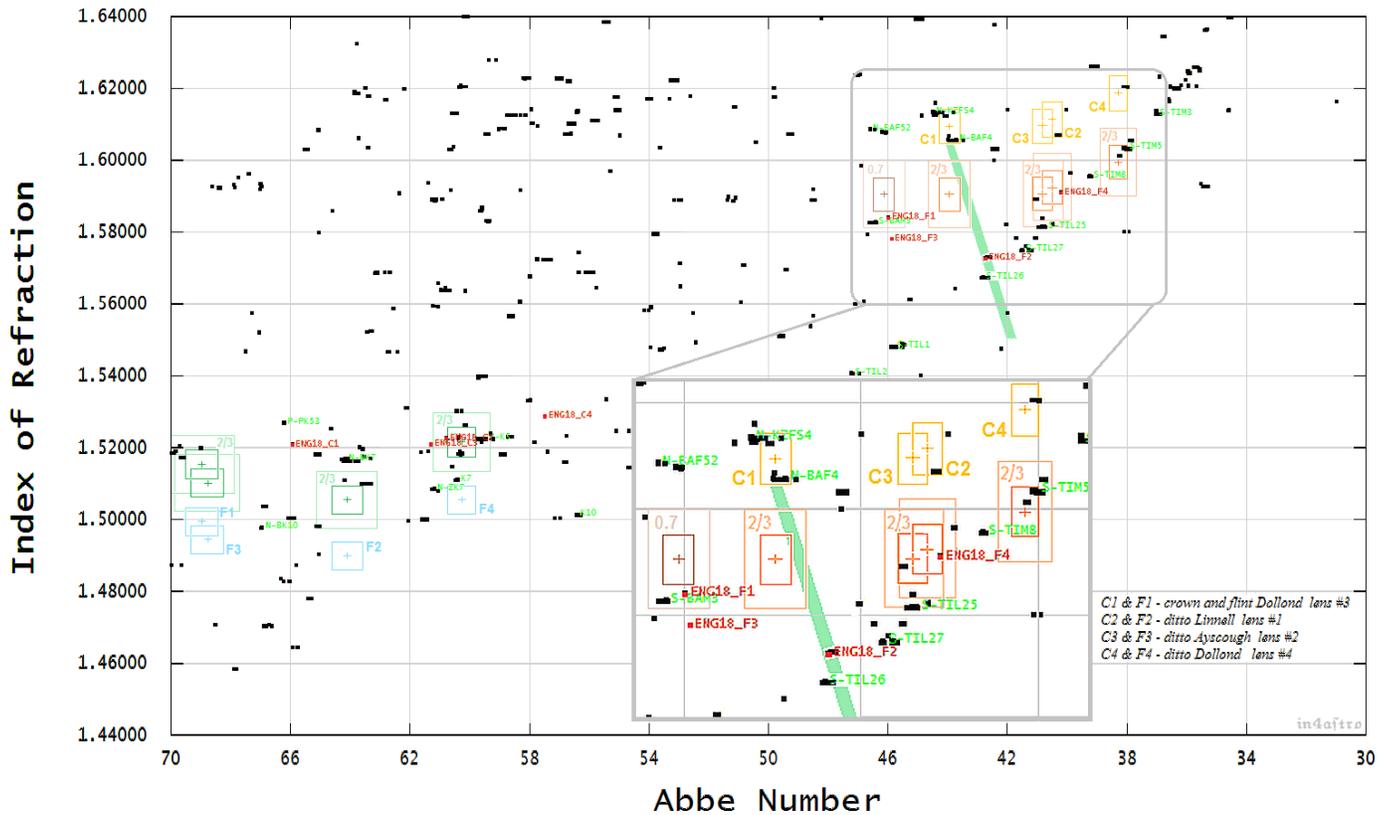

Figure 9. Map of modern glasses (black rectangles) on the diagram "refraction index - Abbe number" from catalogues: Schott, Ohara, USSR and English glasses of the XVIII century (marked in red). The colored rectangles show the regions of correspondence for crowns and flints (details in the text). The light green stripe limits the glasses corresponding to experiment #1.

By the time of experimenting with glass prisms, most likely J. Dollond had changed a technique of observation. Direct visual observation of any objects through prisms did not allow to achieve required accuracy for measurements of angles, achromatic ratios and etc. Probably J. Dollond turned to the Newton's method: in a dark room through a small aperture (about 10mm) in opaque curtains Sunlight was put on and then reflected by a flat mirror in horizontal direction to prisms with input aperture about 0.5mm. Thus, a pinhole camera with the investigated dispersive element was obtained. On this way, instead of direct visual observations could possible to observe the Sun image on a screen, which can be placed on a large distance for increasing sensitivity of this method.

The Fig.10 shows the above described possible scheme of observation with two glass prisms made from F1 flint and C1 crown (glasses from achromatic doublet #3). The flint prism with apex angle about 25.4° and crown prism with apex angle about 28.6° requires to compensate the refraction. The edges of the Sun image are strongly colored (Fig. 10A). The crown prism with apex angle 39.5°±0.1° is required for chromaticism compensation of the flint wedge. This dual prism has the total refractive angle about 7.8° (Fig.10B). In this case, the ratio of the refractive angles of the flint to crown prisms is 0.697. At the prisms–screen distance about 3m, the accuracy of observation scheme should be about 0.1° (Fig.10C).

If we choose other pair of glasses: F4 flint and C2 crown, the acquired pictures qualitatively do not differ from the previous glass pair. In this case, the differences have only quantitative character: for a flint prism with apex angle about 25.3° the crown prism with apex angle about 28.7° is required to compensate refraction; for chromatism compensation a crown prism with apex angle 41.5°±0.1° is required, thus the total angle of refraction of both prisms is about 9.6°. And finally, the ratio of achromatic refracting angles for the flint and crown prisms is 0.651.

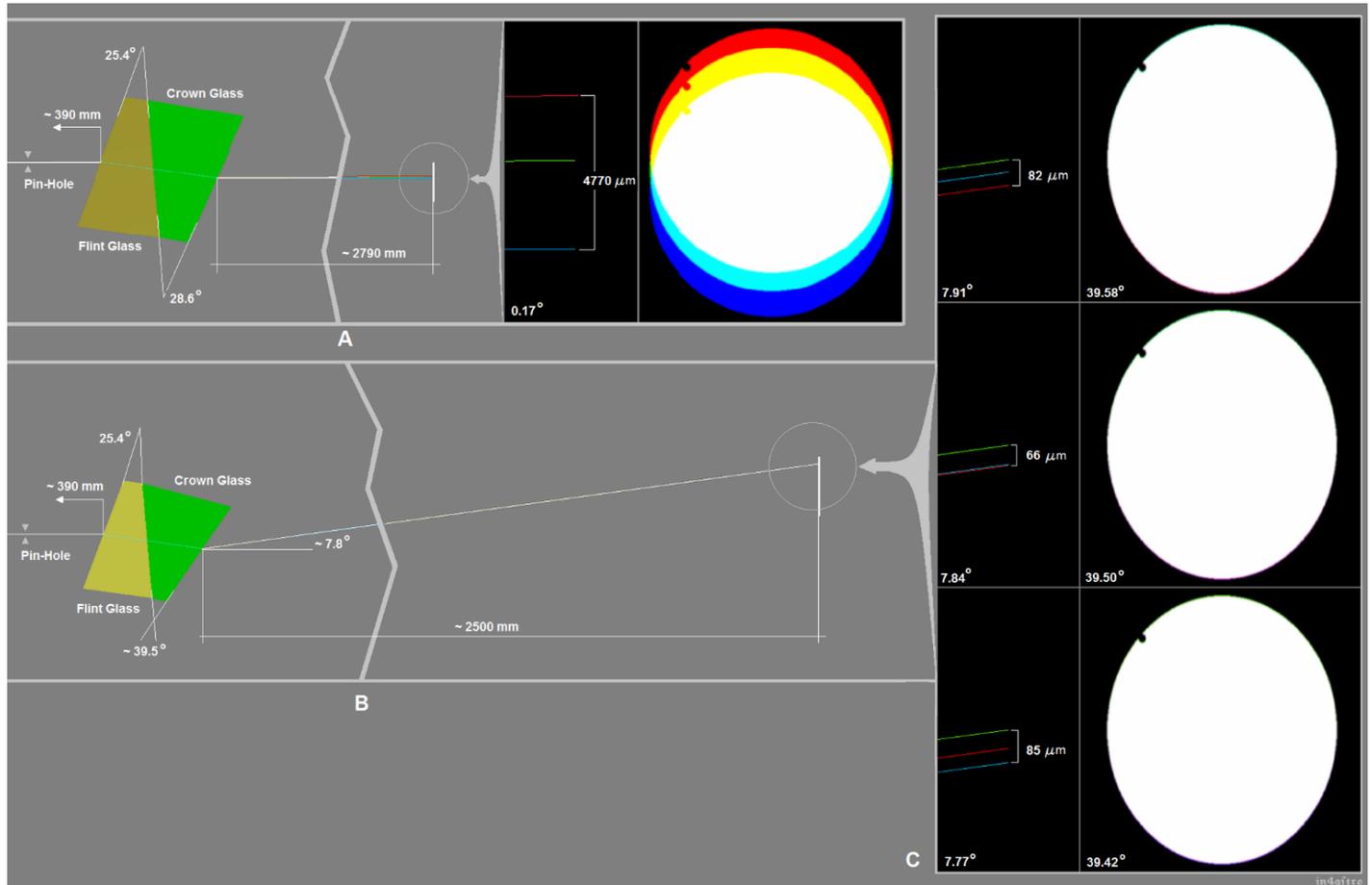

Figure 10. Optical scheme of the observation with two opposing oriented prisms from the F1 flint and C1 crown. The Figure 10A is the optical scheme in the case of reciprocal compensation of refracting angles at the prisms. The dispersion and residual refractive angle (in the lower left corner) and the Sun image are shown. Figure 10B shows the optical scheme in the case of compensated chromatism. Figure 10C shows images of the Sun for the different apex angles (indicated at lower left corner) of crown prism, and next to them, the dispersion of the three rays and the total refractive angle for each case are shown.

## Conclusion

*But the difficulties attending the practice are very considerable. In the first place, the focal distances, as well as the particular surfaces, must be very nicely proportioned to the densities or refracting powers of the glasses; which are very apt to vary in the same sort of glass made at different times. Secondly, the centres of the two glasses must be placed truly on the common axis of the telescope, otherwise the desired effect will be in great measure destroyed. Add to these, that there are four surfaces to be wrought perfectly spherical...* p.742 [1]

In general, simulation of three experiments showed that the old glasses of XVIII century in the achromatic lenses described in work [3] correspond to the results of measurements from the account of J. Dollond [1]. Some discrepancies are minor and can be explained at reasonable assumptions: about observation methods and parameters of used apparatus, about possible accuracies of measurements with subsequent rounding of results when rewriting in the account. There is only one essential mistake in the account – this is statement about "double dispersion" of the water-glass prism to the water prism with the same apex angle. It can be attributed to a misunderstanding that arose in a process of transferring the experimental results from a workbook to the account.

As a result, it is possible to prescribe conditions for selecting achromatic pairs of modern glasses most close by optical parameters to glasses which could be used by J. Dollond in his experiments, and J Dollond & Son optical shop in production of achromatic refractors.

These should be:
- The crown (regular **K** or borosilicate **BK** are preferable; phosphate **PK** are acceptable, but not desirable) with refractive index $n_d \approx 1.52 \div 1.53$ and Abbe number $\nu_d \approx 66 \div 58$
- The flint (preferably classic type with lead oxide **LLF** or **LF**; with barite oxide **BAF** is acceptable, but not desirable) with refractive index $n_d \approx 1.57 \div 1.59$ and Abbe number $\nu_d \approx 46 \div 40$.

For example, there may be such a crown–flint glass pairs:

| Crown (manufacture) | $n_d$ | $\nu_d$ | Flint (manufacture) | $n_d$ | $\nu_d$ |
|---|---|---|---|---|---|
| S-NSL5(Ohara) | 1.5225 | 59.8 | S-TIL25(Ohara) | 1.5814 | 40.7 |
| N-K5(Schott) | | 59.5 | N-LF5(Schott) | | 40.8 |
| S-BSL7(Ohara) | 1.5163 | 64.1 | S-TIL27(Ohara) | 1.5750 | 41.5 |
| N-BK7(Schott) | 1.5168 | 64.2 | | | |
| K18(GOST) | 1.5192 | 60.3 | LF5(GOST) | 1.5750 | 41.3 |
| K19(GOST) | 1.5188 | 61.6 | LF7(GOST) | 1.5784 | 41.1 |

No doubt, this computer simulation should be additionally confirmed by real experiments with one of suggested glass pair. The reconstruction of Dollond's experiments with real prisms could reveal some missing features at their implementation and allow to estimate a labor expenses for them. Measurement of the main parameters of glasses is an important part of the patented method for manufacture of achromatic lenses.

### Acknowledgements

The author is grateful to Yuri Petrunin ([Telescope Engineering Company](), USA) for fruitful discussions and recommendations on the article subject.

### References

[1] J. Dollond, An Account of Some Experiments Concerning the Different Refrangibility of Light, Phil. Trans. Vol.50 (1758), pp. 733-743

[2] D. Jaecks, An investigation of the eighteenth-century achromatic telescope, Annals of Science, 67:2, pp. 149-186

[3] R Willach, New Light on the Invention of the Achromatic Telescope Objective, Notes and Records of the Royal Society of London, Vol.50, No.2 (July 1996), pp. 195-210

[4] А. А. Немиро, Астрономический музей Пулковской обсерватории, в сб. "100 лет Пулковской обсерватории", Пулково, (1939), публ. 1945, стр. 269-271

[5] P. Dollond, Some Account of the Discovery, made by the late John Dollond, F.R S. which led to the grand Improvement of Refracting Telescopes & etc., in book "The Life of John Dollond, F.R.S. inventor of the Achromatic Telescope" by John Kelly, (1808), pp. 61-77

# Appendix I: Optical properties of XVIII century English glasses

Initial information about the glasses is taken from the work [3]. Then, using Conrady formula[16] with a help of Zemax-EE software, parameters of dispersion curves for each glasses were found.

| Glass Name | Refraction index (wavelength, µm) | | | | | | Abbe number[17] | |
|---|---|---|---|---|---|---|---|---|
| | n(0.4861) | n(0.5461) | n(0.5550) | n(0.5893) | n(0.6394) | n(0.6563) | $n_e$ (0.5461) | $n_D$ (0.5893) |
| ENG18_F1 | 1.5930 | 1.5870 | 1.5863 | 1.5839 | 1.5811 | 1.5803 | 46.22 | 45.98 |
| ENG18_F2 | 1.5822 | 1.5759 | 1.5752 | 1.5726 | 1.5696 | 1.5688 | 42.98 | 42.73 |
| ENG18_F3 | 1.5873 | 1.5810 | 1.5803 | 1.5780 | 1.5754 | 1.5747 | 46.09 | 45.85 |
| ENG18_F4 | 1.6013 | 1.5945 | 1.5937 | 1.5909 | 1.5876 | 1.5866 | 40.43 | 40.18 |
| ENG18_C1 | 1.5264 | 1.5227 | 1.5223 | 1.5208 | 1.5190 | 1.5185 | 66.16 | 65.92 |
| ENG18_C2 | 1.5285 | 1.5246 | 1.5241 | 1.5225 | 1.5205 | 1.5199 | 61.00 | 60.76 |
| ENG18_C3 | 1.5272 | 1.5230 | 1.5225 | 1.5209 | 1.5192 | 1.5187 | 61.50 | 61.26 |
| ENG18_C4 | 1.5352 | 1.5310 | 1.5305 | 1.5287 | 1.5266 | 1.5260 | 57.70 | 57.45 |

| Glass name | Conrady formula parameters | | |
|---|---|---|---|
| | $n_0$ | A | B |
| ENG18_F1 | 1.55993 | $1.094782 \cdot 10^{-2}$ | $8.452383 \cdot 10^{-4}$ |
| ENG18_F2 | 1.54652 | $1.218175 \cdot 10^{-2}$ | $8.506154 \cdot 10^{-4}$ |
| ENG18_F3 | 1.56335 | $3.694356 \cdot 10^{-3}$ | $1.309559 \cdot 10^{-3}$ |
| ENG18_F4 | 1.55931 | $1.566757 \cdot 10^{-2}$ | $7.816215 \cdot 10^{-4}$ |
| ENG18_C1 | 1.50500 | $7.474143 \cdot 10^{-3}$ | $4.821722 \cdot 10^{-4}$ |
| ENG18_C2 | 1.50196 | $1.076589 \cdot 10^{-2}$ | $3.518144 \cdot 10^{-4}$ |
| ENG18_C3 | 1.509767 | $3.524723 \cdot 10^{-3}$ | $8.155963 \cdot 10^{-4}$ |
| ENG18_C4 | 1.50750 | $1.095746 \cdot 10^{-2}$ | $4.134970 \cdot 10^{-4}$ |

---

[16] $n(\lambda) = n_0 + A/\lambda + B/\lambda^{3.5}$, here $n_0$, A, B – free parameters are found from three refractive indices which measured at three different wavelengths, usually used d-line (0.5876 µm), F-line (0.4861 µm) and C-line (0.6563 µm).

[17] $\nu(\lambda) = [n(\lambda) - 1]/[n(0.4861) - n(0.6563)]$, $\lambda$ – wavelength in µm, for which the Abbe number is found.